\documentclass[ASNA,onecolumn]{USG}

\usepackage{todonotes}

\usepackage{anyfontsize}
\usepackage{graphicx}
\usepackage{caption}
\usepackage{microtype}

\usepackage[T1]{fontenc}
\usepackage[utf8]{inputenc}
\usepackage[english]{babel}
\graphicspath{{figures/}}
\journal{Color Research \& Application}
\articledoi{}

\begin{document}
\raggedbottom

\title{An Acousto-Optic Colorimeter}
\author[1,2]{P. A. Aksentsev}[][\textsuperscript{*}]
\author[1,2]{G. R. Lobarev}
\author[1,2]{S. A. Korchagin}
\author[1]{O. V. Polschikova}
\author[1,2]{E. I. Ershov}
\authormark{AKSENTSEV \textsc{et al.}}
\titlemark{AN ACOUSTO-OPTIC COLORIMETER}
\address[1]{\orgname{Color Reproduction and Synthesis Institute, }%
\orgaddress{\city{Moscow, }\country{Russia}}}
\address[2]{\orgname{Moscow Center for Advanced Studies, }%
\orgaddress{\city{20, Kulakova Str., Moscow, }\country{Russia}}}

\corres{P. A. Aksentsev (\email{aksentsevpavel@gmail.com})}

\abstract[ABSTRACT]{
Uniform color spaces and color-difference formulae rest on visual data obtained almost entirely with surface samples and conventional displays, whose primaries confine the measurements to a limited chromaticity range. The highly saturated region of color space is therefore the least measured part of human color perception, at the moment when quantum-dot, microLED, and laser displays have begun to operate inside it. Closing this gap requires stimuli more saturated than any fixed-primary device can produce, together with the stability and uniformity demanded by visual experiments. We present an acousto-optic colorimeter built for this purpose. A four-channel acousto-optic tunable filter selects narrow spectral components from a supercontinuum laser over the calibrated range of 443--660~nm, with measured linewidths of 1.37--3.20~nm; the wavelength and radiance of every component are set electronically, so the primary set itself becomes an experimental variable. A compact diffuser--light-pipe homogenizer combines the selected components into a uniform observation field, and spectral feedback stabilizes their intensities. The calibrated primary set covered 95\% of the area enclosed by the CIE 1964 $10^{\circ}$ spectral locus in the $x,y$ chromaticity plane. The calculated maximum luminance reached 1406~cd\,m$^{-2}$, with 328~cd\,m$^{-2}$ available at the gray Robertson chromaticity. Across an 18-mm observation field the mean spatial $\Delta E_{00}$ difference was 0.913, and five Robertson color centers reproduced under continuous feedback gave mean target errors of 0.34--1.26~$\Delta E_{00}$ over 10-min intervals. The instrument thus delivers calibrated, uniform, and stable highly saturated stimuli, and is intended to extend the empirical basis of color-difference measurement into the region where data are still absent.
}

\maketitle

\section{Introduction}\label{sec:introduction}

Human color vision has been measured quantitatively for a century. Color-matching, color-discrimination, and color-difference experiments carried out since the 1920s supplied the data from which the CIE standard observers, the uniform color spaces, and the color-difference formulae still in use were derived \cite{MacAdam1942,Brown1957,Stiles1959,Bongard1956,LuoRigg1986}. These experiments share a constraint that is rarely stated explicitly: their stimuli were surface samples viewed under standard illuminants, or mixtures of display primaries, and both confine the observations to a limited chromaticity range. The empirical basis of colorimetry is consequently dense near the neutral axis and sparse toward the spectral locus, as the coverage of the accumulated visual datasets makes plain \cite{Luo2023}. Where high-chroma data have been gathered deliberately, as in the NCSU-B2 dataset at a high-chroma blue center, the accuracy and the relative ranking of the formulae differ from those established at the less saturated color centers on which formulae were fitted \cite{Shamey2014}. Highly saturated color remains the least explored region of human color perception.

This gap has become a practical one. Quantum-dot, microLED, and laser-based displays now reproduce colors approaching the Rec.~2020 primaries, which lie on the spectral locus itself \cite{Zhu2015,Huang2020Displays,Wang2021MicroLED,ITU2020}. Calibration, gamut mapping, and color-tolerance specification for such devices are performed in uniform color spaces, so every saturated color they render is judged by a metric extrapolated beyond the data that defined it. Evaluations using wide-gamut display colors confirm that the performance of nominally uniform color spaces and color-difference formulae depends on the color center and on the magnitude of the difference \cite{Liang2018,Zhao2020,Xu2021}. Those evaluations are themselves bounded by the display gamut, and the number of primaries and their spectral shapes are fixed by the device: the region in which the metrics are least reliable is precisely the region that fixed-primary hardware cannot reach.

Filling it requires stimuli more saturated than any fixed-primary display can generate, presented in a field that is spatially uniform and radiometrically stable for the duration of a psychophysical session, and with a primary set that can be changed under program control rather than by rebuilding the instrument. The colorimeter described here was built to meet these requirements.

Visual colorimeters present adjustable light mixtures for color-matching and color-discrimination experiments. Such measurements are needed to test color-difference formulas and to characterize observer-dependent matches, particularly for narrowband and highly saturated stimuli \cite{Zhao2020,Xu2021,Asano2020}. Classical instruments used in the work of Brown, Stiles and Burch, and Bongard and Smirnov established the value of spectrally controlled stimuli for these experiments \cite{Brown1957,Stiles1959,Bongard1956}. Electronic control allows calibrated settings to be recalled and stimulus sequences to be programmed, while the achievable gamut and radiometric resolution remain properties of the optical system and its driver.

A visual colorimeter with electronically selectable narrowband primaries controls both the central wavelength and the radiance of each spectral band. Narrow bands enable highly saturated stimuli, and electronic tuning allows the primary set to be changed without replacing emitters. For a target within the gamut of each selected set, the primary radiances can be adjusted to preserve the same XYZ tristimulus values for a specified observer. The resulting spectra are metameric for that observer: their spectral compositions differ, but their calculated colors coincide. This capability supports experiments on observer-dependent color matches and spectral-composition effects. Quantitative use requires characterization of target accuracy, spatial uniformity, and temporal stability.

Programmable sources have used filtered lamps, lasers, and multi-channel LEDs. An early tunable LED colorimetric source combined ten LED groups in an integrating sphere and reported chromaticity stability within 0.001 in $x$ and $y$ during a 50-h optical test, together with source-output stability better than 0.2\% over several hours \cite{Brown2002}. A later vision-research illuminator used six nine-LED heads and a 1-m integrating sphere to provide diffuse spectrally programmable illumination \cite{Mackiewicz2012}. A recent multi-primary visual trichromator reported spectral repeatability as a root-mean-square percentage error (RMSE\%) relative to a mean measured spectrum normalized to unit peak. The short-term values were 0.40 and 0.42 RMSE\% for its two fields, increasing to 0.52 and 0.73 RMSE\% over six months \cite{Shi2024}. Closed-loop control of a 19-channel LED source reduced the target error from more than one to less than $0.4\,\Delta E_{00}$ for the D65-like and A-like illuminants evaluated in that study \cite{Vernet2025}. These results document spectral synthesis and temporal control using different performance measures: changes in optical output, differences between spectra, shifts in chromaticity, or calculated color differences. Each result is therefore quoted in its original metric; the published summary values alone cannot be converted into a common measure of color accuracy.

An acousto-optic tunable filter (AOTF) selects spectral bands by diffracting light from a refractive-index modulation generated by an acoustic wave in a crystal. The radio-frequency (RF) signal driving the acoustic transducer controls the selected wavelength through its frequency and the diffracted intensity through its amplitude \cite{Zhang2018}. AOTFs have been used both for spectral color measurement and for stimulus generation. Zajtsev et al. developed an AOTF-based instrument that calculated color coordinates from measured spectra \cite{Zajtsev2000}. For stimulus generation, AOTFs allow narrow spectral bands to be selected electronically and their relative contributions to be controlled. Machikhin et al. demonstrated wavelength tuning and control of the transmission-band shape in an AOTF-filtered broadband source \cite{Machikhin2018}. Romanova et al. subsequently examined the illumination optics needed to couple a broadband source into the filter \cite{Romanova2021}. Together, these studies addressed both electronic control of the output spectrum and optical coupling into the filter.

Later work applied this approach directly to color reproduction. Beliaeva et al. investigated mixtures of two and three narrow spectral components and the factors affecting their colorimetric accuracy \cite{Beliaeva2022}. Machikhin et al. reproduced 19 ColorChecker targets using AOTF-selected primaries at 460, 520, and 650~nm, reporting mean and median ~$\Delta E_{00}$ values of approximately 1.25 and 1.12~$\Delta E_{00}$ \cite{Machikhin2023}. Color was evaluated from spatially averaged camera data because the unmixed diffracted fields were non-uniform. Optical systems designed specifically for AOTF-based visual colorimetry have also been proposed \cite{Beliaeva2024Visual}.

Previous work has thus established AOTF color synthesis and examined the optical arrangements needed for visual colorimetry. The present study examines how this spectral flexibility can be combined with a sufficiently uniform observation field and reproducible output after calibration. These properties must be evaluated together for an instrument intended for color-vision experiments. The study therefore characterizes spectral linewidths, target accuracy, temporal repeatability, spatial uniformity, and modeled gamut in a single supercontinuum-based instrument, with the observer, reference white, field geometry, and measurement interval specified.

\begin{figure}[!htbp]
\centering
\includegraphics[width=\textwidth]{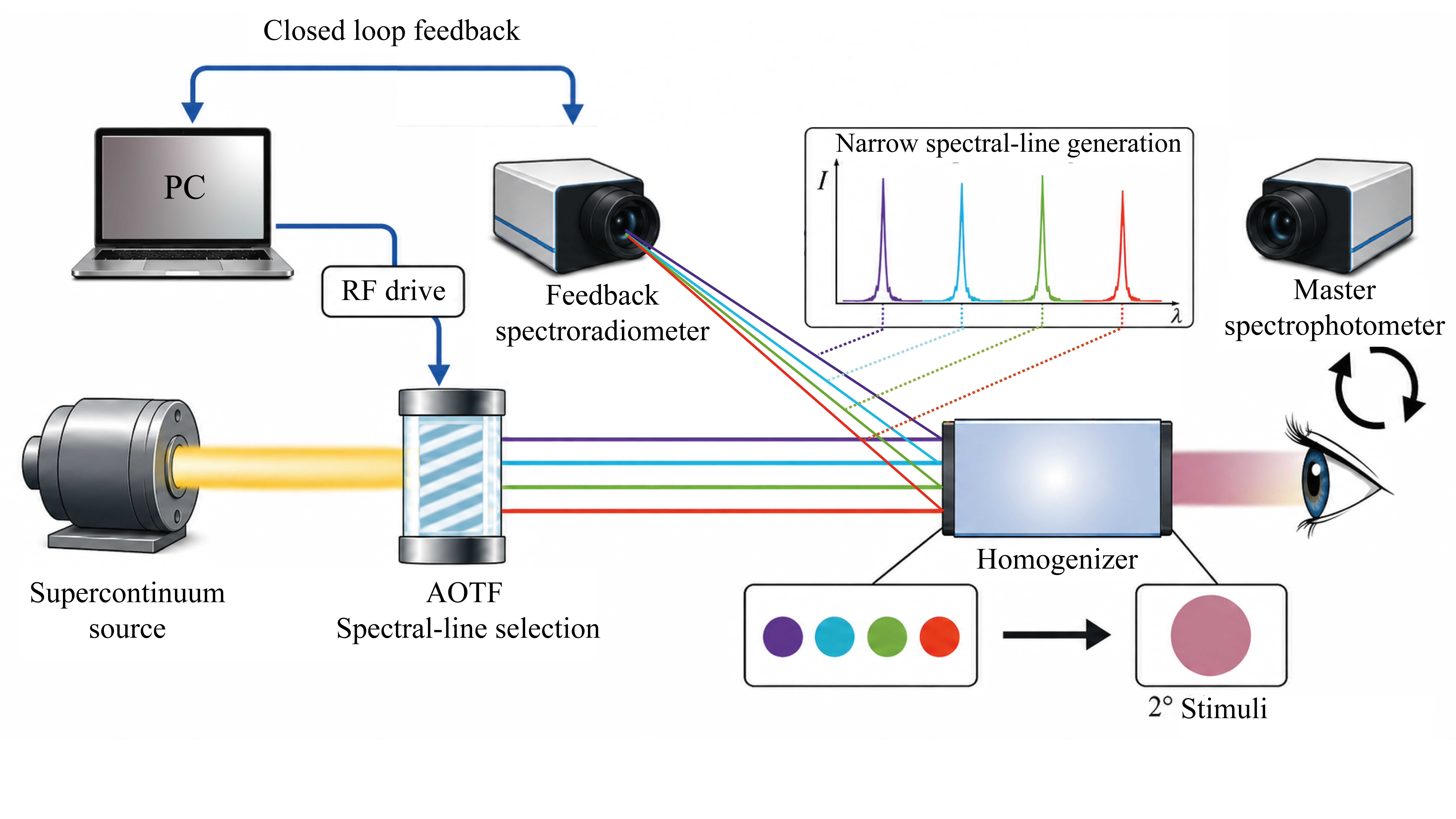}
\caption{Operating principle of the acousto-optic colorimeter. The AOTF filters narrow spectral components from the supercontinuum source, and the homogenizer mixes them to form the observation field. Light reflected at the homogenizer entrance is sampled by the feedback spectroradiometer; the computer adjusts the RF drive to maintain the required output of each component. The field is intended for viewing by a human observer and is characterized here using the master spectrophotometer.}
\label{fig:operating-principle}
\end{figure}

This work presents and experimentally characterizes an acousto-optic colorimeter (AOC), illustrated in Figure~\ref{fig:operating-principle}, based on a four-channel AOTF illuminated by a supercontinuum laser. Supercontinuum generation broadens laser radiation through nonlinear optical processes to produce a continuous broadband spectrum \cite{Dudley2006}. After filtering, the available spectrum spans 380--750~nm. The source has a beam-quality factor $M^2<1.2$, close to the ideal Gaussian-beam value of unity. Across the calibrated wavelengths and four channels, the measured full widths at half maximum (FWHM) are 1.37--3.20~nm. A compact two-stage diffuser--light-pipe homogenizer combines the angularly separated beams, a rotating diffuse screen suppresses stationary speckle by temporal averaging \cite{Shin2006}, and feedback adjusts the intensity of each selected spectral component at 50~Hz. The colorimeter is intended to present controlled color stimuli to human observers. In this study, its output is characterized instrumentally using an X-Rite i1Pro 2 master spectrophotometer in emission mode. Color differences calculated from the measured spectra describe the agreement with the target color and the consistency of the field; no observer judgments are reported here.

The combination of electronically tunable few-nanometer lines and broad chromaticity coverage is the principal motivation for the AOTF architecture. Changing the acoustic drive frequencies selects a different set of primary wavelengths without replacing any emitters. The calibration data are then used to adjust the contribution of each selected line so that their mixture reproduces the same target $X,Y,Z$ values, provided that the target remains achievable with the new primaries. The instrument can change the spectral composition of a stimulus while preserving its calculated color for the chosen standard observer. Multi-primary trichromators have used spectrally selectable stimuli to estimate individual color-matching functions and cone spectral sensitivities \cite{Shi2024}; related work shows that narrow-primary stimuli are also informative for mapping observer metamerism and testing observer-specific color reproduction \cite{Asano2020,Ko2023,Trumpy2023}. The AOC can therefore support controlled experiments on individual color matching, observer metamerism, and color-difference behavior near saturated color centers, provided that its spatial and temporal performance is qualified.

The study describes a calibration-based three-channel color-reproduction algorithm and characterizes the 18-mm observation field using normalized spectral error (NSE), ~$\Delta E_{00}$ color differences $\Delta E_{00}$, and differences $\Delta E'$ in the uniform color space derived from the CAM16 color-appearance model (CAM16-UCS). Five reproduced color centers are measured with the master instrument while feedback remains continuously enabled. The modeled gamut is expressed as the area reachable by the calibrated AOC in the International Commission on Illumination (CIE) 1964 $10^{\circ}$ $x,y$ chromaticity diagram, as a percentage of the area enclosed by the spectral locus and line of purples. A complementary three-primary calculation maps the maximum attainable luminance across this diagram under the power limit used in the calculation.

The hardware can operate with up to four simultaneous spectral channels. Color-reproduction and spatial-uniformity measurements use three active channels, whereas the spectral characterization examines all four channels individually. For three linearly independent primaries with fixed spectral shapes, an achievable target has a unique set of non-negative channel radiances. A fourth primary introduces an additional degree of freedom in the spectral mixture and therefore requires another selection criterion---for example, drive resolution, observer agreement, spectral robustness, or power efficiency---and separate validation. This multi-objective four-channel problem is outside the scope of the present study. Gray was selected as a demanding test of channel balance because small relative errors between narrowband primaries can produce appreciable ~$\Delta E_{00}$ deviations near grayity. Measurements of red, green, blue, and yellow also assess how target accuracy and its chromaticity and luminance contributions vary between operating points. ~$\Delta E_{00}$ is used for temporal accuracy and repeatability \cite{Sharma2005}; CAM16-UCS $\Delta E'$ additionally characterizes spatial non-uniformity \cite{Li2017}. Values near one $\Delta E_{00}$ are discussed only as a nominal perceptibility region because the threshold depends on color center, luminance, field size, adaptation, and observer \cite{Gladilin2025,Huang2025,ZhaoWei2025}.

\clearpage

\section{Methods}\label{sec:methods}

The experimental procedure combines paired spectral calibration, calibration-based synthesis of target colors, and closed-loop regulation of the selected spectral components. Instrument performance is assessed through spectral-linewidth measurements, spatially resolved colorimetry, and repeated measurements at the center of the observation field. Chromaticity coverage is calculated separately from the measured calibration spectra.

\subsection{Optical architecture and spatial homogenization}\label{sec:optical-setup}

\begin{figure}[htbp]
\centering
\includegraphics[width=\textwidth]{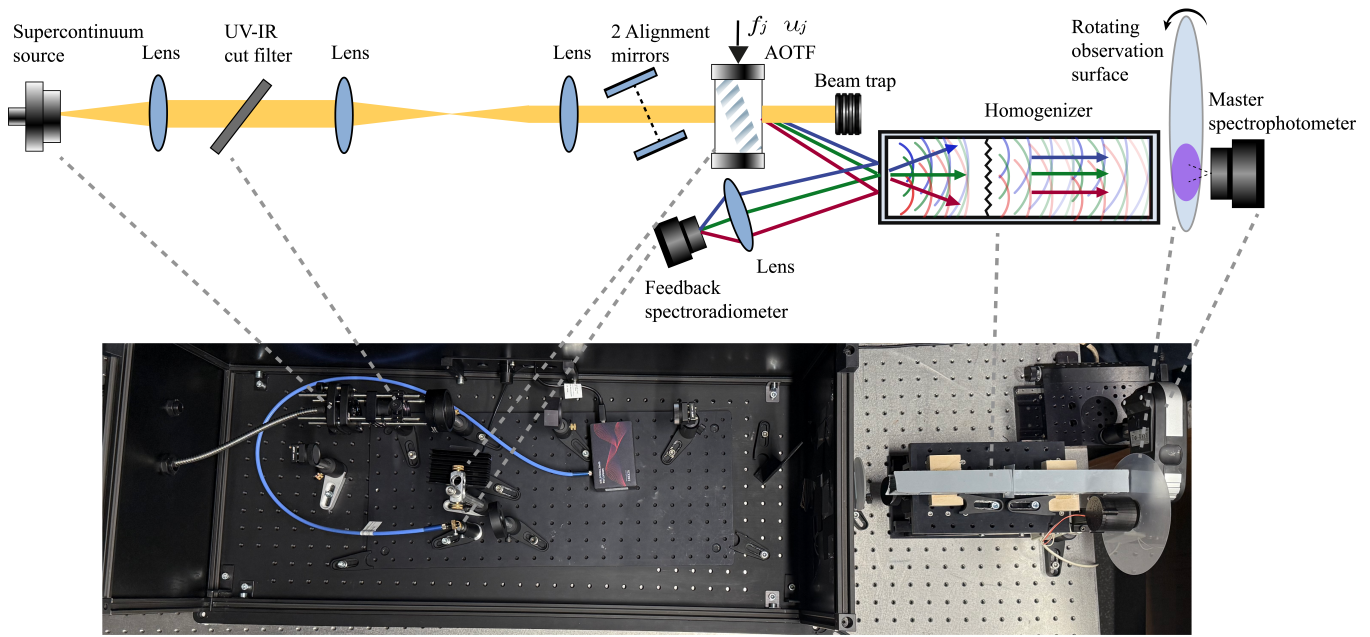}
\caption{Optical layout and photograph of the assembled acousto-optic colorimeter. Dashed gray lines link selected elements in the schematic to their counterparts in the photograph, including the source, UV--IR cut filter, AOTF, feedback spectroradiometer, homogenizer, rotating observation surface, and master spectrophotometer.}
\label{fig:optical-setup}
\end{figure}

The colorimeter shown in Figure~\ref{fig:optical-setup} is based on a four-channel AOTF with a TeO$_2$ acousto-optic cell driven at 70--135~MHz. The first lens collimates the supercontinuum radiation emerging from the output fiber of the SCLS-PS-UV source. Ultraviolet (UV) and infrared (IR) cut filters restrict the transmitted spectrum to 380--750~nm, and a telescope adjusts the beam diameter to provide the required illumination across the AOTF aperture. Steering mirrors provide precise angular and positional alignment of the incident beam. The undiffracted zeroth-order beam is intercepted by a beam stop, whereas the first-order diffracted components are used for color reproduction. The laser beam-quality factor is $M^2<1.2$. Across the calibrated wavelengths and four channels, the measured FWHM was 1.37--3.20~nm; the measurement procedure is described in Section~\ref{sec:spectral-characterization}. For channel $j$, the RF drive frequency $f_j$ sets the selected wavelength $\lambda_j$, whereas the RF amplitude command $u_j$ controls the diffracted line radiance. The pairs $(f_j,u_j)$ are independently specified through the computer-controlled driver.

Unlike the fiber-coupled halogen lamp used by Machikhin et al. \cite{Machikhin2023}, the present system uses a supercontinuum laser with $M^2<1.2$. Its beam can be collimated with low angular divergence, so the light enters the TeO$_2$ cell over a comparatively narrow range of incidence angles. For a fixed incidence geometry and RF frequency, efficient diffraction occurs near the wavelength that satisfies the Bragg phase-matching condition \cite{Zhang2018}. Reducing the spread of incidence angles therefore reduces angular broadening of the selected spectral band and helps preserve wavelength-selective diffraction. This physical advantage follows from the illumination geometry; diffraction efficiency was not compared directly with the lamp-based system. Figure~\ref{fig:optical-setup} combines the optical layout and a photograph of the assembled instrument; dashed gray lines connect corresponding components.

The AOTF diffracts different wavelengths at different angles, producing spatially separated beams that require homogenization. The combination of diffusing screens and repeated reflections inside a mirror-lined rectangular light pipe is established in visual colorimetry \cite{Cavonius1974}. The present two-stage diffuser--light-pipe homogenizer uses this combination to mix the spectral components and form the required observation field. The 18-mm field subtended approximately $2^{\circ}$ at 515~mm; this geometric viewing angle is distinct from the CIE 1964 $10^{\circ}$ color-matching functions used for the numerical colorimetric calculations reported below.

The high spatial and temporal coherence of the isolated spectral lines produces a granular speckle pattern when the light is scattered by the observation surface. The diffusive surface is therefore rotated continuously, causing the speckle realization to change throughout the finite integration time of an observer or detector. The resulting temporal average strongly suppresses the stationary granular structure \cite{Shin2006}.

Two instruments are used for characterization and control. The X-Rite i1Pro 2 spectrophotometer, operated in emission mode, serves as the master instrument and records the absolute spectral radiance used by the color-reproduction algorithm described in Section~\ref{sec:color-reproduction}. The YIXIST YOA-8404-05 serves as the feedback spectroradiometer, with a nominal spectral sampling interval of 0.3~nm per detector pixel. The feedback algorithm uses its wavelength-resolved detector signal in relative units.

\subsection{Calibration and radiometric monitoring}\label{sec:calibration}

The SCLS-PS-UV supercontinuum laser was allowed to stabilize for 2~h after switch-on before calibration. Its filtered output covered 380--750~nm and had a beam-quality factor $M^2<1.2$. Calibration was then performed for each physical AOTF channel $j$ and selected radio frequency $f_j$, which sets the line center $\lambda_j$. The RF amplitude command was swept through its usable range in 1\% increments, with $k$ indexing the measured command levels $u_{j,k}$ of channel $j$. At each level, both instruments recorded a wavelength-resolved spectrum. After the reference normalization defined below, each calibration-table entry is the paired record
\begin{equation}
\mathcal C_{j,k}=
\left\{f_j,u_{j,k},L_j(\lambda;u_{j,k}),R_j(\lambda;u_{j,k})\right\},
\label{eq:calibration-record}
\end{equation}

where $L_j(\lambda;u_{j,k})$ is the spectral radiance measured by the master spectrophotometer in emission mode and $R_j(\lambda;u_{j,k})$ is the simultaneous relative spectral response of the feedback spectroradiometer. For each selected line, the same wavelength interval $W_j$ is used for both instruments in reference normalization and line-level monitoring:
\begin{equation}
W_j=\left[\lambda_j-\frac{\Delta\lambda_j}{2},\lambda_j+\frac{\Delta\lambda_j}{2}\right],
\qquad
\Delta\lambda_j=3\times10^{-7}\lambda_j^3.
\label{eq:integration-window}
\end{equation}
Here, $\lambda_j$ and the empirically selected integration width $\Delta\lambda_j$ are expressed in nanometers. The interval is fixed for each selected line. These window integrals provide reference and feedback values; colorimetric calculations use the full measured spectra.

The present dataset contains 19 frequency settings at 3-MHz intervals. This sampling interval is an experimental choice rather than an intrinsic limit. The RF synthesizer has a frequency-setting resolution of 0.01~MHz over 70--135~MHz, so a denser wavelength calibration can be acquired with smaller RF-frequency increments. Acquisition of the full amplitude response takes approximately 2~min per selected line and channel. The duration is governed primarily by the X-Rite i1Pro~2 acquisition time, which the instrument selects automatically for each spectrum.

During the calibration scan, a reference stimulus with fixed RF amplitude command $r$ was recorded with the X-Rite i1Pro~2 every 5~min at each selected reference frequency $f_\ell$, where $\ell$ indexes reference frequencies. Let $h$ index successive reference acquisitions at that frequency, and let $L_{\ell,h}^{\mathrm{ref}}(\lambda;r)$ denote the measured reference spectral radiance. Its integral $c_{\ell,h}^{\mathrm{ref}}$ is evaluated over $W_\ell$, the interval defined by Equation~\eqref{eq:integration-window} at the corresponding reference wavelength. After the scan, the mean reference integral $\overline{c}_{\ell}^{\mathrm{ref}}$ is calculated separately for each frequency, and the associated calibration records are rescaled to that mean:
\begin{equation}
\begin{aligned}
c_{\ell,h}^{\mathrm{ref}}&=\int_{W_\ell}L_{\ell,h}^{\mathrm{ref}}(\lambda;r)\,\mathrm d\lambda,
&\overline{c}_{\ell}^{\mathrm{ref}}&=\frac{1}{H_\ell}\sum_{h=1}^{H_\ell}c_{\ell,h}^{\mathrm{ref}},\\
q_{\ell,h}&=\frac{\overline{c}_{\ell}^{\mathrm{ref}}}{c_{\ell,h}^{\mathrm{ref}}},
&L_j&=q_{\ell,h}L_j^{\mathrm{raw}},\qquad R_j=q_{\ell,h}R_j^{\mathrm{raw}}.
\end{aligned}
\label{eq:reference-normalization}
\end{equation}
Here, $H_\ell$ is the number of X-Rite reference acquisitions at frequency $f_\ell$, and $q_{\ell,h}$ is the normalization factor. The superscript $\mathrm{raw}$ denotes spectra before reference normalization; $L_j$ and $R_j$ denote the resulting normalized master and feedback spectra. Their wavelength and command arguments are omitted in Equation~\eqref{eq:reference-normalization} for brevity. The same factor is applied to both spectra of each associated calibration record, preserving their correspondence while compensating for slow source variation.

\subsection{Three-channel color-reproduction algorithm}\label{sec:color-reproduction}

 Let $\Lambda=\{\lambda_j:j=1,2,3\} $ denote calibrated wavelengths. For each primary, the integrated line radiance $c_j$ serves as a calibration-table coordinate. The table provides the corresponding RF amplitude command $u_j(c_j)$ and measured spectral radiance $L_j(\lambda;u_j(c_j))$. The synthesized spectrum is the sum of these primary spectra:
\begin{equation}
L_{\mathrm{mix}}(\lambda)=
\sum_{j} L_j(\lambda;u_j(c_j)).
\end{equation}
\label{eq:mixture-spectrum}

For a trial vector $\mathbf c$ of channel radiances, the columns of the primary matrix are the tristimulus values calculated from the full corresponding spectra:
\begin{equation}
\mathbf A_\Lambda(\mathbf c)=
\left[
\begin{array}{c}
X_j\\Y_j\\Z_j
\end{array}
\right]_{j},
\qquad
\begin{bmatrix}X_j\\Y_j\\Z_j\end{bmatrix}
=
\kappa\int
L_j(\lambda;u_j(c_j))
\begin{bmatrix}
\bar{x}_{10}(\lambda)\\
\bar{y}_{10}(\lambda)\\
\bar{z}_{10}(\lambda)
\end{bmatrix}\,\mathrm d\lambda .
\label{eq:primary-matrix}
\end{equation}
The integral extends over the full recorded spectral range, rather than the line-monitoring window $W_j$. The functions $\bar{x}_{10}$, $\bar{y}_{10}$, and $\bar{z}_{10}$ are the CIE 1964 $10^{\circ}$ color-matching functions. The factor $\kappa$ converts the recorded spectral-radiance units to the tristimulus scale, with $Y$ expressed in $\mathrm{cd\,m^{-2}}$.

The target is specified by the tristimulus vector $\mathbf t=[X_{\mathrm t},Y_{\mathrm t},Z_{\mathrm t}]^{\mathrm T}$. At a fixed trial point, a nonsingular $\mathbf A_\Lambda(\mathbf c)$ gives the dimensionless correction factors $\mathbf a$:
\begin{equation}
\mathbf a=\mathbf A_\Lambda(\mathbf c)^{-1}\mathbf t,
\qquad
c_j^{\mathrm{new}}=a_jc_j .
\label{eq:three-channel-solution}
\end{equation}
Thus, $a_j$ specifies the multiplicative correction to the trial radiance of channel $j$.

For each trial $c_j$, the algorithm identifies the two adjacent 1\%-spaced commands whose measured master line radiances bracket that value and linearly interpolates the paired spectra $L_j$ and $R_j$. The interpolated master spectra form $\mathbf A_\Lambda(\mathbf c)$, Equation~\eqref{eq:three-channel-solution} updates the trial radiances, and the table lookup is repeated until the radiances converge. This iteration accounts for the small changes in line shape with command level. The converged command $u_j$ is rounded to the 0.1\% hardware command step. Interpolation is confined to measured command levels at each calibrated wavelength; wavelengths themselves are not interpolated. A wavelength set is accepted only when all requested radiances are non-negative, remain within the calibrated channel ranges, and satisfy the shared electrical-power limit of the RF driver. Candidate wavelength triplets are ranked by the predicted color displacement caused by one 0.1\% command step. The Gray operating point used in the present measurements was reproduced with $\Lambda=\{457,480,573\}~\mathrm{nm}$.

\subsection{Closed-loop stabilization}\label{sec:feedback-loop}

The feedback loop regulates the integrated signal of each selected spectral line by adjusting the same RF amplitude command $u_j$ used by the color-reproduction algorithm. For each paired calibration record, the master line radiance and the feedback line integral are
\[
\begin{aligned}
c_j(u_{j,k})&=\int_{W_j}L_j(\lambda;u_{j,k})\,\mathrm d\lambda,\\
m_j(u_{j,k})&=\int_{W_j}R_j(\lambda;u_{j,k})\,\mathrm d\lambda.
\end{aligned}
\]
Thus, the calibration table contains the two line integrals measured at the same command.

To compensate for the optical-power drift of the supercontinuum source, the light reflected from the uncoated first surface of the homogenizer is measured by the YIXIST YOA-8404-05 feedback spectroradiometer. Because the AOTF drive frequency is highly stable, the spectral-peak positions are obtained from the preliminary calibration.

For each active channel, the feedback spectrum is integrated over the interval $W_j$ defined in Equation~\eqref{eq:integration-window}. The color-reproduction solution supplies the required master-instrument line radiance $c_j^\star$. Through the paired calibration record, this value determines the command $u_j^\star$ and the corresponding target line value $m_j^\star$ of the feedback spectroradiometer. The current line value is obtained from the spectrum measured at update $n$:
\begin{equation}
\begin{aligned}
m_j^\star
&=\int_{W_j}R_j(\lambda;u_j^\star)\,\mathrm d\lambda,\\m_j[n]\vphantom{m_j^\star}
&=\int_{W_j}R_j(\lambda;n)\,\mathrm d\lambda.
\end{aligned}
\label{eq:monitor-line-values}
\end{equation}
Thus, $R_j$ is the wavelength-resolved feedback-spectroradiometer signal, whereas $m_j^\star$ and $m_j[n]$ are the target and current line integrals used by the controller. The feedback law adjusts the current feedback value to match the calibration value paired with the master-instrument radiance required for color reproduction. Although three channels were active in the present experiment, the same procedure is defined for each of the four AOTF channels.

The controller is updated at $f_{\mathrm{upd}}=50~\mathrm{Hz}$. At update $n$, the current line value $m_j[n]$ is compared with the target $m_j^\star$. The command increment is proportional to the line-value error:
\begin{equation}
\begin{aligned}
e_j[n] &= m_j^\star-m_j[n],\\
u_j[n+1] &= u_j[n]+K_{\mathrm p}e_j[n],
\end{aligned}
\label{eq:signal-error}
\end{equation}
where $e_j[n]$ is the line-value error and $K_{\mathrm p}$ is the proportional gain. The initial command $u_j[0]=u_j^\star$ is taken from the converged calibration-table solution. The 50-Hz value specifies the digital update frequency, not the closed-loop bandwidth, which was not measured.

The feedback loop was continuously enabled throughout the central-color measurements used to quantify accuracy and repeatability. The RF driver supports amplitude-command updates at up to 1~kHz; this hardware setting rate is distinct from the 50-Hz feedback update frequency and does not specify the optical settling time.

\subsection{Drive-range characterization and equalization}\label{sec:apparat-accuracy-increasing}

The output amplitude of the four-channel RF driver decreases monotonically over the 70--135-MHz operating range, with an amplitude variation of approximately one order of magnitude between the lowest and highest frequencies. The Bragg diffraction efficiency of the AOTF is also highly non-uniform across the tuning range: it exhibits distinct maxima near the center of the optical tuning range and in its long-wavelength region, with a pronounced minimum in the short-wavelength region. The combined wavelength dependence of the RF-driver output and diffraction efficiency therefore produces a strongly non-uniform optical response.

\begin{figure}[htbp]
\centering
\includegraphics[width=\textwidth]{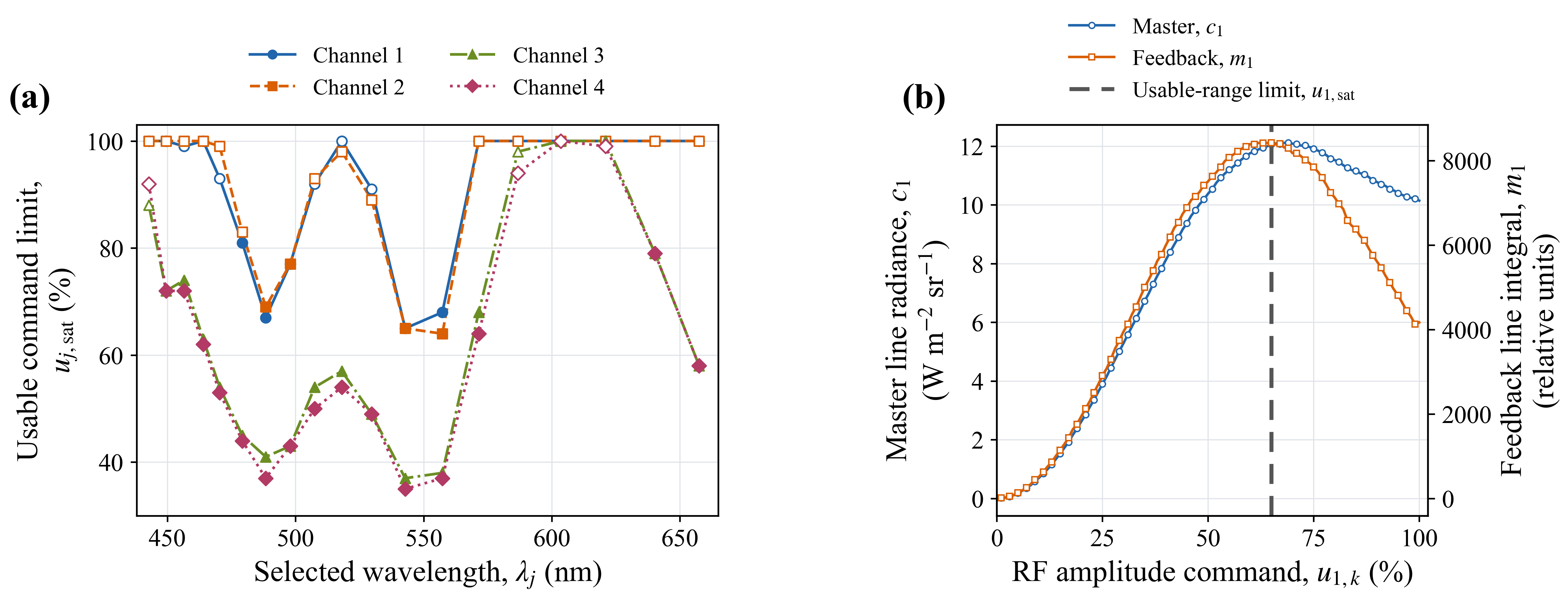}
\caption{Drive-range characterization. (a) Saturation or measured-range command limit $u_{j,\mathrm{sat}}$ versus selected wavelength $\lambda_j$ for the four AOTF channels under the attenuation configuration used in this study. Filled markers denote saturation detected by the five-drop criterion; open markers denote the measured-range peak when that criterion was not triggered. (b) Channel-1 master line radiance $c_1$ and feedback line integral $m_1$ versus RF amplitude command at approximately 543~nm, illustrating the response decrease beyond saturation. The vertical dashed line marks the usable command limit.}
\label{fig:saturation-limits}
\end{figure}

The nonlinear command--radiance response depends on the acousto-optic material and the acoustic field established by the particular cell design \cite{Mi2025}. The saturation behavior must therefore be characterized for each channel and wavelength. Once a spectral channel is driven beyond saturation, the spectral-peak amplitude decreases and the spectral line broadens. Commands in this regime were excluded because they reduce the available peak line radiance and degrade spectral selectivity. The usable command limit $u_{j,\mathrm{sat}}$ was determined from each amplitude scan. Saturation was identified when five consecutive values each showed a relative reduction of at least 10\% from the preceding peak; if this criterion was not triggered, the command at the measured maximum was retained as the range limit. For a nominal 10-bit full scale, the corresponding number of usable command levels is $1024u_{j,\mathrm{sat}}/100$. Across the four channels and 19 wavelength settings, $u_{j,\mathrm{sat}}$ ranged from 35\% to 100\%, corresponding to approximately 358--1024 command levels [Figure~\ref{fig:saturation-limits}(a)]. Figure~\ref{fig:saturation-limits}(b) illustrates the rise and subsequent decline of the measured line integrals with increasing command at a representative wavelength.

To distribute the available command levels more uniformly across the spectral range, selective attenuation is applied to the channels (Figure~\ref{fig:saturation-limits}). The channel allocation can be reassigned as required by the experimental conditions.

\subsection{Characterization procedures}\label{sec:characterization-procedures}
The characterization procedure was designed to evaluate the main properties of the AOC relevant to controlled color-stimulus generation. The measurements addressed three complementary aspects of instrument performance: spectral characteristics of the generated primaries, spatial uniformity of the observation field, and color-reproduction accuracy and stability. In addition, the calibrated spectral data were used to estimate the accessible chromaticity region and the maximum attainable luminance across it.

\subsubsection{Spectral characterization}\label{sec:spectral-characterization}

Spectral linewidths were evaluated from the YIXIST spectra for all four AOTF channels and 19 RF settings (Figure~\ref{fig:fwhm}(a)). For each channel--frequency pair, the spectrum recorded at the saturation or measured-range peak command defined in Section~\ref{sec:apparat-accuracy-increasing} was analyzed over 410--750~nm. The peak was located, the two half-maximum crossings were found by linear interpolation relative to a zero baseline, and the measured FWHM $\Delta\lambda_{\mathrm{FWHM},j}$ was calculated as their wavelength separation. All widths were calculated on the wavelength grids stored with the measured spectra.

An auxiliary channel-1 scan at a fixed 30\% RF amplitude command provided the wavelength--frequency relation shown in Figure~\ref{fig:fwhm}(b). The representative channel-1 spectra illustrating the half-maximum calculation in Figure~\ref{fig:fwhm}(c) were recorded at a fixed 60\% command.

\subsubsection{Spatial uniformity}

Spatial uniformity of the resulting color field was evaluated over a circular 18-mm-diameter observation region image using hyperspectral data from the accompanying homogenizer study. The image data were dark-corrected and converted to a radiance scale using the central X-Rite measurement and the per-line spectral-response correction established for the two instruments. Each pixel spectrum was compared with the mean spectrum of the region. Tristimulus values were calculated over 400--730~nm for the CIE 1964 $10^{\circ}$ standard observer and converted using a D50 reference white. The analysis quantified NSE, $\Delta E_{00}$, and CAM16-UCS $\Delta E'$. To separate spectral-shape variation from luminance variation, NSE was calculated both from the radiance spectra and after normalizing each pixel spectrum to the same spectral integral.

\subsubsection{Color-reproduction accuracy and stability}

Color-reproduction accuracy and short-term repeatability were evaluated using measurements from the central region of the homogenized field viewed by the master spectrophotometer in emission mode. The five targets correspond to the CIE reference color centers specified in Robertson's guidelines for color-difference evaluation \cite{Robertson1978}. Their target coordinates are given in Table~\ref{tab:color-errors}. The tabulated $Y$ values are the absolute luminances used for these emitted stimuli, in $\mathrm{cd\,m^{-2}}$. During measurements of color-difference steps around each target, the central color was measured repeatedly; these central-color readings form the accuracy series analyzed here. Feedback remained continuously enabled during acquisition. The stability measurement experiment began 12~h after calibration. Immediately before feedback was enabled, the calibration-based gray setting gave an error of approximately $3\,\Delta E_{00}$. Each measured stimulus was converted to the CIE 1976 $L^*a^*b^*$ color space (CIELAB) using the CIE 1964 $10^{\circ}$ standard observer and a D50 reference white at the 100~$\mathrm{cd\,m^{-2}}$ adaptation level. ~$\Delta E_{00}$ $\Delta E_{00}$ was calculated with unit parametric factors \cite{Sharma2005}. X-Rite specifies short-term repeatability of $0.1\,\Delta E_{94}$ on white under its stated D50, $2^{\circ}$ test conditions \cite{XriteI1Pro2}; this is an instrument repeatability specification rather than a general uncertainty bound for the present narrowband measurements.

To assess chromaticity and luminance mismatch separately, $\Delta E_{00}$ was recalculated for each central-color reading with the measured $x_{10},y_{10}$ retained and $Y$ set to the target value, and then with the measured $Y$ retained and $x_{10},y_{10}$ set to the target coordinates. The total error and the two fixed-coordinate errors were averaged separately over the readings included in each color series. Each result is the mean of the individual target errors, rather than an error calculated from an averaged stimulus. All comparisons used the same D50 reference white and unit parametric factors; the two fixed-coordinate differences are not additive.

\subsubsection{Chromaticity-gamut and luminance calculation}

The chromaticity gamut was calculated from the 19 calibrated wavelengths independently of the stability measurements. The calibrated primary spectra were converted to CIE 1964 $10^{\circ}$ tristimulus values and then to $x,y$ chromaticities. Each three-primary set defines a triangle of chromaticity mixtures in the linear non-negative model; the union of these triangles defines the modeled AOC region $\Omega_{\mathrm{AOC}}$. For the same primary chromaticities, this union gives the same geometric region as the union of all four-primary convex hulls. The chromaticity coverage was calculated as
\begin{equation}
G_{xy}=100\,\frac{\operatorname{area}(\Omega_{\mathrm{AOC}})}
{\operatorname{area}(\Omega_{\mathrm{SL}})},
\label{eq:gamut-coverage}
\end{equation}
where $\Omega_{\mathrm{SL}}$ is the region enclosed by the CIE 1964 $10^{\circ}$ spectral locus and the line of purples. Thus, $G_{xy}$ is a two-dimensional chromaticity-area ratio, not a three-dimensional volume in CIELAB or CAM16-UCS. This geometric projection does not impose a fixed luminance or the shared electrical-power constraint of the RF driver.

To characterize the attainable luminance, a separate three-primary calculation used the same calibration data to determine $Y_{\max}(x_{10},y_{10})$, the greatest luminance available at each evaluated chromaticity under the power limit imposed in the calculation. The resulting map combines the chromaticity boundary with the luminance attainable within it. For a prescribed luminance $Y$, a chromaticity must additionally satisfy $Y\leq Y_{\max}(x_{10},y_{10})$; chromaticity-area coverage alone does not establish this condition. The map is a calibration-based prediction, not a set of independent measurements at every plotted point.

\section{Results}\label{sec:results}

The results summarize the measured spectral linewidths, field uniformity, accuracy and short-term variation of five reproduced color centers. The principal specifications are followed by the individual measurements and the chromaticity coverage and maximum-luminance map calculated from the calibrated device model.

\subsection{Principal AOC characteristics}

Table~\ref{tab:aoc-characteristics} summarizes the principal optical, control, and measured performance characteristics, distinguishing the hardware capabilities from the three-channel configuration used for the reported measurements.

\begin{table}[htbp]
\centering
\caption{Principal characteristics of the acousto-optic colorimeter.}
\label{tab:aoc-characteristics}
\small
\begin{tabularx}{\textwidth}{@{}p{0.31\textwidth}p{0.27\textwidth}X@{}}
\toprule
Characteristic & Value & Basis or condition \\
\midrule
Optical source & SCLS-PS-UV supercontinuum laser & Filtered output 380--750~nm; $M^2<1.2$ \\
AOTF cell material & TeO$_2$ & Four-channel acousto-optic cell \\
AOTF RF operating range & 70--135~MHz & Multi-frequency drive \\
Simultaneous spectral channels & Up to 4; 3 used here & Independent $(f_j,u_j)$ control \\
\addlinespace[2pt]
Calibrated wavelength grid & 443--660~nm; 19 positions & Current 3-MHz sampling interval; denser calibration is supported \\
Calibrated three-wavelength sets & 969 & Sets of three distinct wavelengths selected from 19; order disregarded \\
RF-frequency setting resolution & 0.01~MHz & Hardware setting resolution over 70--135~MHz \\
RF amplitude command step & 0.1\% (approximately 10 bit) & After channel-dependent attenuation \\
Nominal three-channel command-state upper bound & $1.04\times10^{12}$ & $969\times1024^3$; configurations, not perceptually distinct colors \\
Feedback-spectroradiometer sampling interval & 0.3~nm per pixel & Nominal spectral sampling of the YOA-8404-05 unit \\
Measured spectral-line FWHM & 1.37--3.20~nm & 76 spectra; half-maximum crossings interpolated on the measured wavelength grid \\
Observed command-limit range & 35--100\% (approximately 358--1024 levels) & Four channels and 19 wavelength settings \\
\addlinespace[2pt]
Calibration command increment & 1\% & Interpolation between adjacent measured levels \\
Laser warm-up before calibration & 2~h & Stationary operating state \\
Calibration time & Approximately 2~min per line and channel & Full response; X-Rite acquisition time selected automatically \\
Calibration-to-test interval & 12~h & Accuracy and stability measurements \\
\addlinespace[2pt]
Master/feedback instruments & X-Rite i1Pro 2/YIXIST YOA-8404-05 & Emission-mode spectral radiance/relative closed-loop signal \\
RF amplitude-command update rate & Up to 1~kHz & Driver setting rate; distinct from feedback updates \\
Feedback update frequency & 50~Hz & Line-integral regulation \\
Feedback state during color series & Continuously enabled & No open-loop interval during acquisition \\
Central-color sampling intervals & 5.3~s; 3.4~s for yellow & Master-instrument acquisition cadence \\
\addlinespace[2pt]
Uniformity observation region & 18-mm diameter ($2^{\circ}$ at 515~mm) & Entire formed color field \\
Spatial color difference, $\Delta E_{00}$ & Mean 0.913; 95th percentile 2.126 & Relative to the field-mean spectrum over the 18-mm region \\
Accuracy measurement region & Central field region & Master-spectrophotometer measurement in emission mode \\
\addlinespace[2pt]
Chromaticity diagram coverage, $G_{xy}$ & 95\% & For CIE 1964 $x_{10}y_{10}$ observer; Equation~\eqref{eq:gamut-coverage}; not a fixed-luminance gamut \\
Largest computed luminance & $1406~\mathrm{cd \cdot m^{-2}}$ & Three primaries, power limit applied; $(x_{10},y_{10})=(0.390,0.580)$; Figure~\ref{fig:gamut-luminance} \\
Computed $Y_{\max}$ at the gray Robertson center & $328~\mathrm{cd \cdot m^{-2}}$ & Three primaries, power limit applied; $(x_{10},y_{10})=(0.314,0.331)$; Figure~\ref{fig:gamut-luminance} \\
\bottomrule
\end{tabularx}
\end{table}

The 19 calibrated wavelengths yield 969 sets of three distinct wavelengths, with each set counted once regardless of order ($N_{\Lambda}=969$). If each selected channel could use all nominal 10-bit amplitude codes, the combinatorial upper bound would be $N_{\Lambda} \cdot 1024^3\approx1.04\times10^{12}$ command configurations. This count does not represent perceptually distinct colors: different configurations can be colorimetrically equivalent, and the measured command ceilings, shared RF-driver power limit, and feasibility constraints reduce the usable set.

\begin{figure}[htbp]
\centering
\includegraphics[width=\textwidth]{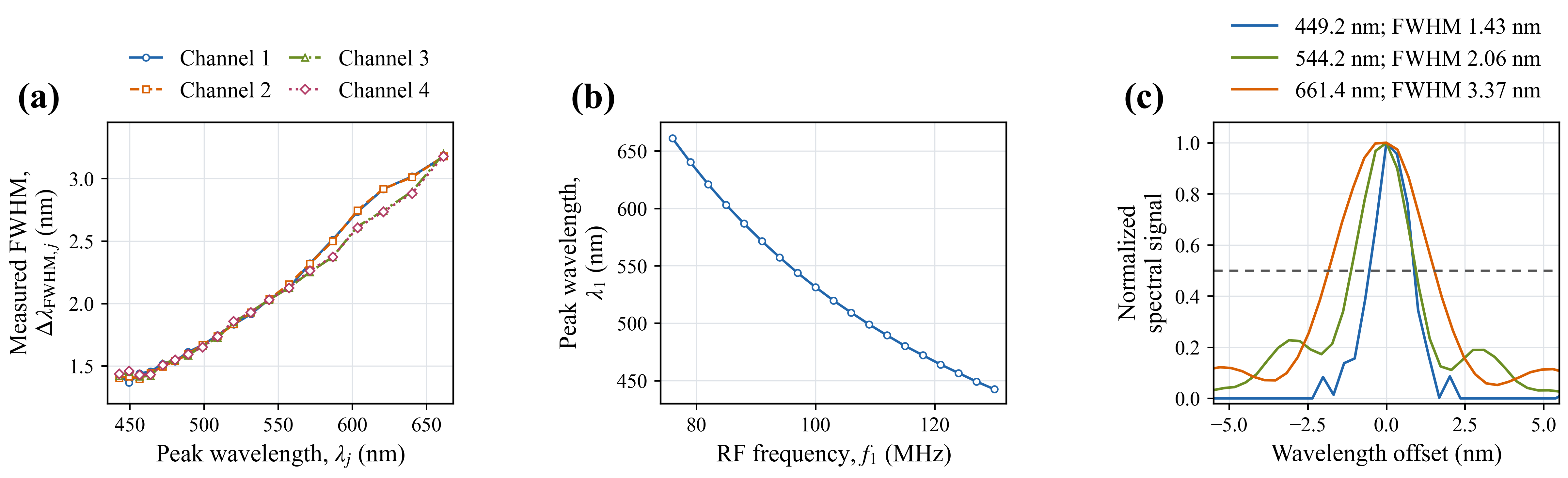}
\caption{Spectral characterization with the YIXIST YOA-8404-05 feedback spectroradiometer. (a) Measured full width at half maximum $\Delta\lambda_{\mathrm{FWHM},j}$ versus spectral-peak wavelength $\lambda_j$ for all four AOTF channels, evaluated at the saturation or measured-range peak command. (b) Channel-1 spectral-peak wavelength versus RF frequency at a fixed 30\% command. (c) Representative channel-1 spectra at a fixed 60\% command, normalized to their respective maxima and centered on their measured peaks at 449.2, 544.2, and 661.4~nm; markers denote recorded spectral samples, and the dashed line marks the half-maximum level.}
\label{fig:fwhm}

\end{figure}

Figure~\ref{fig:fwhm} summarizes three spectral measurements. Figure~\ref{fig:fwhm}(a) shows the FWHM values obtained from the 76 spectra recorded at the commands corresponding to the usable-range peaks. Across the four channels, $\Delta\lambda_{\mathrm{FWHM},j}$ ranged from 1.37 to 3.20~nm, with an overall mean of 2.00~nm. The nearly coincident channel trends indicate a common wavelength dependence, with broader lines toward the red end of the calibrated range. At a fixed 30\% command in channel~1, the spectral peak shifted monotonically from 661.1~nm at 76~MHz to 442.4~nm at 130~MHz (Figure~\ref{fig:fwhm}(b)). Representative short-, middle-, and long-wavelength spectra recorded at a fixed 60\% command (Figure~\ref{fig:fwhm}(c)) had FWHM values of 1.43, 2.06, and 3.37~nm at peak wavelengths of 449.2, 544.2, and 661.4~nm, respectively. These three spectra illustrate the half-maximum calculation at a common command; their widths are reported separately from the 76-spectrum summary obtained at the usable-range peak commands.

\subsection{Spatial uniformity}
\begin{figure}[htbp]
\centering
\includegraphics[width=\textwidth]{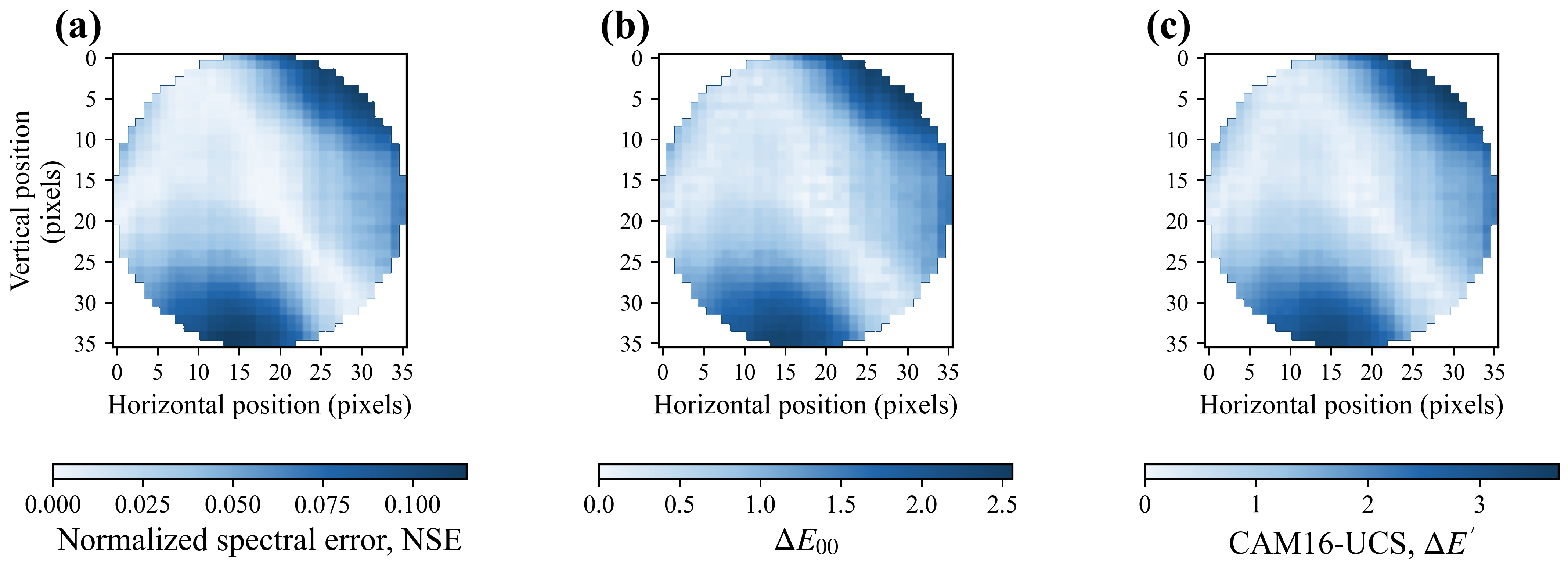}
\caption{Spatial error maps over the 18-mm-diameter observation region for the bright stimulus with field-mean coordinates $x=0.2686$, $y=0.3206$, and $Y=42.7~\mathrm{cd\,m^{-2}}$: (a) NSE, (b) $\Delta E_{00}$, and (c) CAM16-UCS $\Delta E'$. The radiance-scaled spectrum at each pixel is compared with the mean spectrum of the region. Darker blue indicates a larger error; Figures~\ref{fig:field-error-maps}(a)--(c) use separate numerical scales.}
\label{fig:field-error-maps}
\end{figure}

For the bright test stimulus, the mean color coordinates obtained from the hyperspectral field data were $x=0.2686$, $y=0.3206$, and $Y=42.7~\mathrm{cd\,m^{-2}}$. The residual non-uniformity arose mainly from variation in absolute radiance across the field. The normalized spectral shapes were substantially more uniform, indicating effective mixing of the selected wavelengths despite the uneven luminance distribution. The NSE had a mean of 0.040, a median of 0.033, a 95th percentile of 0.097, and a maximum of 0.115. Normalizing each pixel spectrum to the same spectral integral reduced these values to 0.012, 0.010, 0.025, and 0.033, respectively. The 70\% reduction in mean NSE identifies luminance variation as the dominant contribution to the residual field error; the non-zero normalized NSE quantifies the smaller remaining variation in spectral shape. The mean, median, 95th-percentile, and maximum ~$\Delta E_{00}$ differences were 0.913, 0.753, 2.126, and 2.561, respectively; the corresponding CAM16-UCS $\Delta E'$ values were 1.263, 1.047, 3.042, and 3.706. The maps in Figure~\ref{fig:field-error-maps} show smooth, large-scale variation rather than predominantly granular fine structure, consistent with a residual radiometric gradient after spectral mixing.

\subsection{Color-reproduction accuracy and stability}

\begin{figure}[!htbp]
\centering
\includegraphics[width=\textwidth]{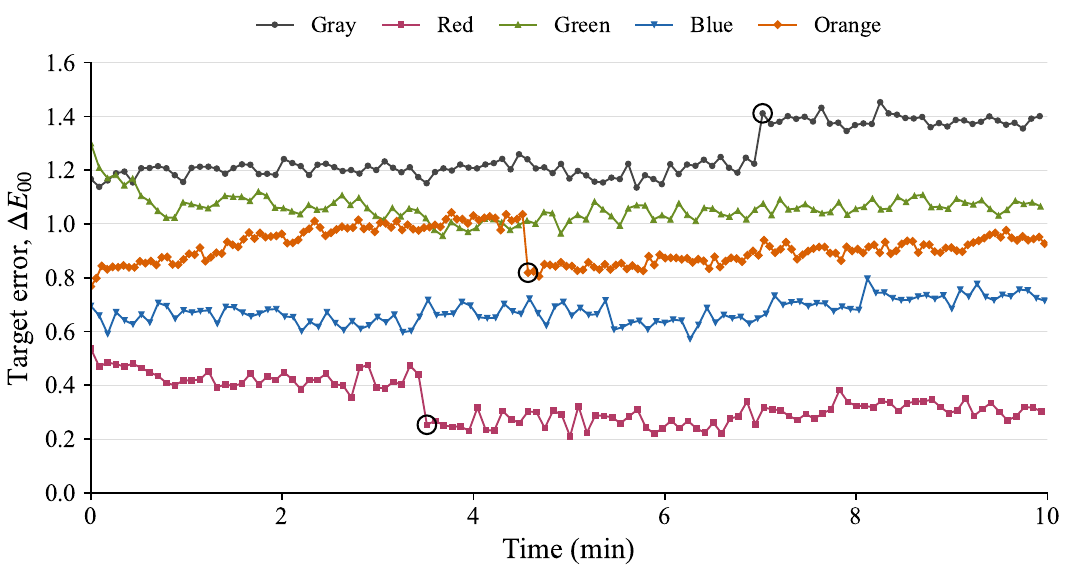}
\caption{Target color error $\Delta E_{00}$ for sequential central-field color measurements with feedback continuously enabled. Time coordinates are interpolated within acquisition passes. Open circles mark the abrupt changes in the gray, red, and yellow series discussed in the text. The means in Table~\ref{tab:color-errors} include all displayed measurements, including these changes.}
\label{fig:white-stability}
\end{figure}

Figure~\ref{fig:white-stability} shows the central-field color measurements for gray, red, green, blue, and yellow over 10~min per color. The mean interval between successive master-instrument measurements was 5.3~s for gray, red, green, and blue, and 3.4~s for yellow. This cadence reflects spectral acquisition during the stimulus sequence, including the integration time selected automatically by the master instrument; it is not the rate at which the driver can change the stimulus. All measurements within each window are shown, including the abrupt changes observed for gray, red, and yellow. Mean target errors were 1.26, 0.34, 1.06, 0.67, and 0.91~$\Delta E_{00}$, respectively. The corresponding standard deviations were 0.09, 0.08, 0.05, 0.04, and 0.06~$\Delta E_{00}$. These values describe the complete displayed series, including changes in level, rather than fluctuations about a single stationary level.

Table~\ref{tab:color-errors} lists the target coordinates, selected primary wavelengths, and mean total, chromaticity-only, and luminance-only errors for the same measurements. The chromaticity-only error was larger for gray, green, blue, and yellow, whereas the luminance-only error was larger for red. Their relative magnitudes thus depend on the reproduced color and cannot be inferred from the gray-point result alone.

\begin{table}[htbp]
\centering
\caption{Mean central-field color errors for the five Robertson reference centers \cite{Robertson1978} over the same 10-min windows as Figure~\ref{fig:white-stability}. Target coordinates use the CIE 1964 $10^{\circ}$ observer, with $Y$ in $\mathrm{cd\,m^{-2}}$. Primary wavelengths correspond to physical channels 2, 3, and 4, respectively. The total, chromaticity-only, and luminance-only errors are expressed as $\Delta E_{00}$. The chromaticity-only and luminance-only comparisons defined in Section~\ref{sec:characterization-procedures} are non-additive.}
\label{tab:color-errors}
\fontsize{9}{11}\selectfont
\setlength{\tabcolsep}{7pt}
\renewcommand{\arraystretch}{1.2}
\begin{tabular}{@{}lllrrr@{}}
\toprule
& & & \multicolumn{3}{c}{Mean $\Delta E_{00}$} \\
\cmidrule(l){4-6}
Color & Target $(x_{10},y_{10},Y)$ & Wavelengths (nm) & Total & Chromaticity & Luminance \\
\midrule
Gray & $(0.314,0.331,30.0)$ & 480, 457, 573 & 1.26 & 1.20 & 0.24 \\
Red & $(0.484,0.342,14.1)$ & 500, 473, 640 & 0.34 & 0.19 & 0.27 \\
Green & $(0.248,0.362,24.0)$ & 573, 463, 490 & 1.06 & 0.90 & 0.52 \\
Blue & $(0.219,0.216,8.8)$ & 573, 450, 473 & 0.67 & 0.62 & 0.17 \\
Yellow & $(0.388,0.428,69.3)$ & 490, 573, 450 & 0.91 & 0.89 & 0.29 \\
\bottomrule
\end{tabular}
\end{table}

The marked level changes show the colorimetric effect of the minimum single-channel drive adjustment. Changing one channel by one RF amplitude command increment changes its optical power and, consequently, the reproduced color. The $\Delta E_{00}$ differences between the stimuli immediately before and after the marked transitions were 0.22 for gray, 0.32 for red, and 0.23 for yellow.

\subsection{Chromaticity gamut and attainable luminance}

\begin{figure}[!htbp]
\centering
\includegraphics[width=0.88\textwidth]{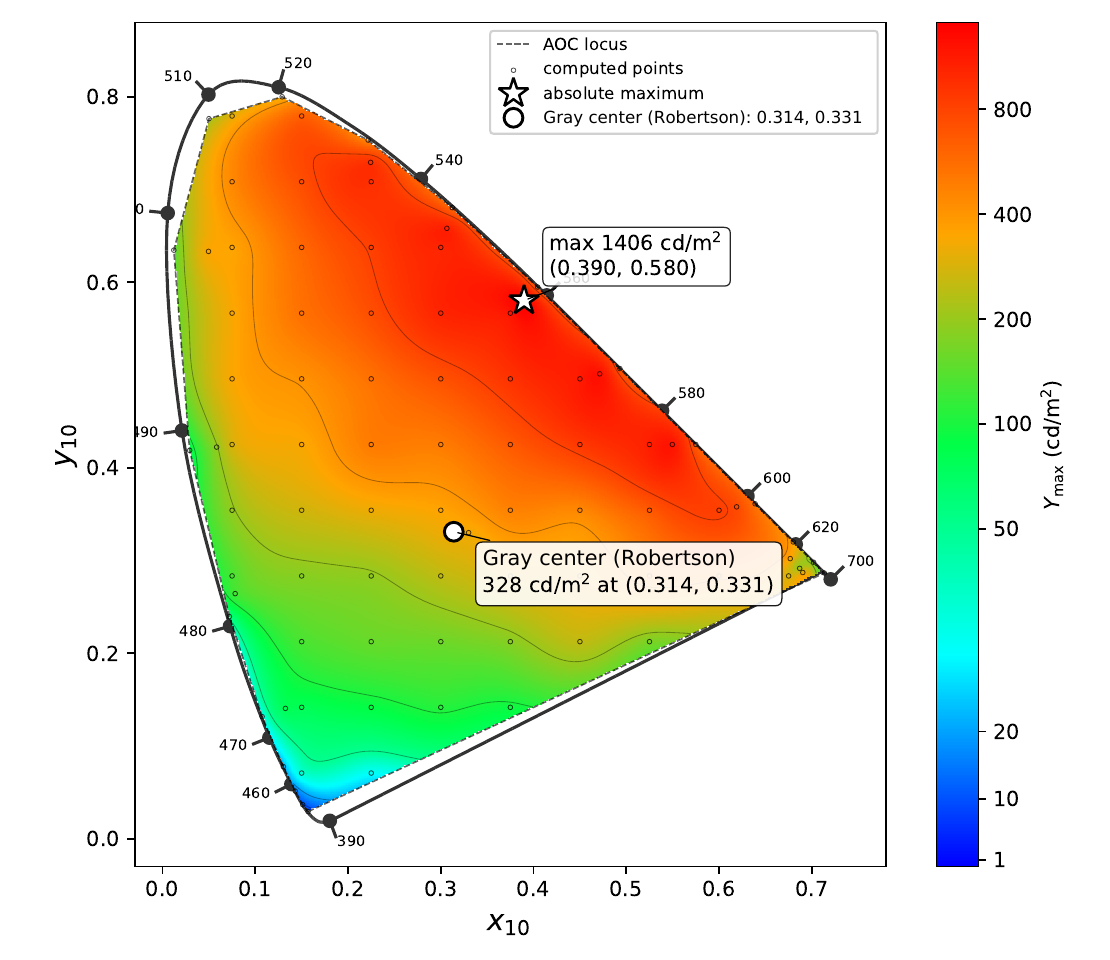}
\caption{Heat map of modeled xyY gamut in three-primary operation, calculated from the current 19-wavelength calibration with the power limit applied. The color denotes the maximum achievable luminance in the xy point. The dashed curve marks the AOC chromaticity boundary; the solid outer curve is the CIE 1964 $10^{\circ}$ spectral locus, labeled by wavelength in nanometers. Small open circles denote the evaluated chromaticities, and the false-color scale and contours represent $Y_{\max}$ in $\mathrm{cd \cdot m^{-2}}$. The star marks the largest computed value, $1406~\mathrm{cd \cdot m^{-2}}$ at $(0.390,0.580)$. The large open circle marks the gray Robertson chromaticity, for which the calculated limit is $328~\mathrm{cd \cdot m^{-2}}$.}
\label{fig:gamut-luminance}
\end{figure}

Applying Equation~\eqref{eq:gamut-coverage} to the calibrated primary set gave $G_{xy}=95\%$. Figure~\ref{fig:gamut-luminance} shows the corresponding three-primary map of $Y_{\max}$ under the power limit used in the calculation. The largest computed luminance was $1406~\mathrm{cd\,m^{-2}}$ at $(x_{10},y_{10})=(0.390,0.580)$; at the gray Robertson chromaticity $(0.314,0.331)$, $Y_{\max}=328~\mathrm{cd\,m^{-2}}$. The latter is a calculated output limit, not the $30~\mathrm{cd\,m^{-2}}$ target used for the gray series in Table~\ref{tab:color-errors}. Maximum luminance varies substantially with chromaticity, reaching its highest values in the yellow--green region and lower values in the short-wavelength region. The false-color scale represents $Y_{\max}$, not the appearance of the generated stimuli.

For context, a review of display technologies reports 91.8\% Rec.~2020 coverage for an OLED implementation and 91.4\% for a red--green--blue (RGB) microLED implementation \cite{Huang2020Displays}. In the CIE 1931 $x,y$ diagram, the Rec.~2020 primary triangle occupies 63.37\% of the area enclosed by the spectral locus and line of purples; the reported coverages therefore correspond to approximately 58.2\% and 57.9\% of that area, respectively. This conversion uses the Rec.~2020 primary coordinates \cite{ITU2020} and a polygonal area calculation from the CIE 1931 color-matching functions at 1-nm intervals over 360--830~nm. The present $G_{xy}=95\%$ uses the CIE 1964 $10^{\circ}$ observer, so the percentages provide context rather than a strictly common-observer comparison. The AOC area ratio describes geometric chromaticity coverage, while Figure~\ref{fig:gamut-luminance} additionally shows the luminance constraint. Neither constitutes spectroradiometric verification of every boundary point.

\section{Discussion}\label{sec:discussion}

The measured color errors and field non-uniformity are examined in relation to the optical response and feedback control, then compared with published results. Potential improvements are considered through calibration, primary selection, and optical equalization.

\subsection{Measured accuracy, repeatability, and field uniformity}

In the five central-color series acquired with continuous feedback, mean target errors were 0.34--1.26~$\Delta E_{00}$. Gray had the largest mean error among the tested settings, consistent with its role as a demanding test of narrowband-primary balance. Errors near one ~$\Delta E_{00}$ unit are often compared with a nominal perceptibility threshold, but measured thresholds depend on color center, luminance, field size, adaptation, observer, and experimental method \cite{Gladilin2025,Huang2025,ZhaoWei2025}. Visual detectability was not measured here, so these instrumental errors do not establish perceptual indistinguishability.

Twelve hours after calibration, the gray-point error was approximately $3\,\Delta E_{00}$ immediately before feedback was enabled, compared with a mean of $1.26\,\Delta E_{00}$ during subsequent regulated operation. This difference supports the use of feedback to correct the calibration-based setting. The overall standard deviations in Figure~\ref{fig:white-stability} include changes in operating level and therefore do not represent stationary noise alone. The manufacturer's $0.1\,\Delta E_{94}$ gray-repeatability specification applies to a different metric and test condition and is not an uncertainty estimate for these measurements. Likewise, the 50\nobreakdash-Hz controller update frequency does not establish the closed-loop bandwidth, and the central-color acquisitions do not resolve the controller dynamics.

Table~\ref{tab:color-errors} shows that chromaticity mismatch is the larger contributor to target error for gray, green, blue, and yellow. Uniformly rescaling their output radiance would therefore leave most of the mismatch uncorrected; improving accuracy requires a better relative balance of the selected lines. For red, the luminance-only error is larger: 0.27 versus $0.19\,\Delta E_{00}$ for chromaticity alone. These comparisons include measurement variability and describe target accuracy, not the separate contributions to temporal variance.

Feedback regulates radiometric quantities rather than $\Delta E_{00}$ directly. For fixed primary spectra, the line radiances $c_j$ contribute linearly to the synthesized spectrum and XYZ tristimulus values, and the calibrated feedback line values $m_j$ vary monotonically with these radiances. Although CIELAB conversion and ~$\Delta E_{00}$ are nonlinear, the effects of small radiance perturbations near the calibrated operating point can be approximated to first order in color coordinates. This local relationship motivates regulation of the individual line integrals but does not imply uniform gain or colorimetric sensitivity throughout the gamut.

The mean spatial $\Delta E_{00}$ was below one unit, although the 95th-percentile and maximum values reached 2.126 and 2.561, respectively. The mean NSE decreased from 0.040 to 0.012 after equal-integral normalization, indicating that luminance variation dominates the residual field non-uniformity. Homogenizer refinement should therefore focus on flattening the radiometric field and improving alignment while preserving spectral mixing.

\subsection{Comparison with published programmable sources}

Table~\ref{tab:published-comparison} compares field uniformity, temporal or target performance, and spectral and gamut characteristics. The original metrics and conditions are retained for each study because the published summary values cannot be converted to a common performance measure without the underlying data.

\begin{table}[htbp]
\centering
\caption{Published performance data relevant to programmable color-stimulus sources. Values retain the metrics and test conditions of the cited studies and do not constitute a direct ranking.}
\label{tab:published-comparison}
\scriptsize
\begin{tabularx}{\textwidth}{@{}>{\raggedright\arraybackslash}p{0.145\textwidth}>{\raggedright\arraybackslash}p{0.205\textwidth}>{\raggedright\arraybackslash}p{0.285\textwidth}>{\raggedright\arraybackslash}X@{}}
\toprule
System & Field characterization & Temporal or target performance & Spectral/gamut scope \\
\midrule
Brown et al. \cite{Brown2002} & Integrating-sphere output; fieldwise perceptual mapping was outside the reported scope & $x,y$ within 0.001 during a 50-h test; output better than 0.2\% over several hours & Ten LED groups; standard-source synthesis over 380--780~nm; no chromaticity-area ratio \\
\addlinespace[5pt]
Mackiewicz et al. \cite{Mackiewicz2012} & Diffuse output from a 1-m integrating sphere & Linear and repeatable drive response described; no color-difference time series & Six nine-LED heads; spectral and color matching; no single gamut-area value \\
\addlinespace[5pt]
Shi et al. \cite{Shi2024} & Separate $2^{\circ}$ and $10^{\circ}$ fields; fieldwise error mapping was outside the reported scope & Spectral RMSE\%: short-term 0.40/0.42; six-month 0.52/0.73 & 18 LEDs per field, 400--700~nm; 11 primary triplets plotted in CIE 1976 $u'v'$ without one coverage percentage \\
\addlinespace[5pt]
Vernet et al. \cite{Vernet2025} & Fieldwise uniformity was outside the stabilization study & D65-like and A-like targets: $>1\,\Delta E_{00}$ without stabilization and $<0.4\,\Delta E_{00}$ with stabilization & 19 selected visible channels; target-spectrum synthesis; no gamut-area metric \\
\addlinespace[5pt]
Machikhin et al. \cite{Machikhin2023} & Unmixed fields were spatially averaged for color evaluation & Mean/median target errors: 1.25/1.12~$\Delta E_{00}$ for 19 ColorChecker targets & Three AOTF bands; near-full CIE 1931 coverage described without a directly comparable area denominator \\
\addlinespace[5pt]
Yuan et al. \cite{Yuan2013} & Planar radiometric uniformity $>99.5\%$ over a central 890-mm region of a 1-m exit port & Two-hour radiometric fluctuation $<0.12\%$ & Spectrally tunable source, 400--2500~nm; radiometric spectral matching rather than a perceptual gamut \\
\addlinespace[5pt]
Present AOC & Compact diffuser--light-pipe homogenizer; mean/p95/max $\Delta E_{00}=0.913/2.126/2.561$ over 18~mm & Five color centers with feedback enabled: per-color mean errors 0.34--1.26 $\Delta E_{00}$ over 10~min each; complete-series standard deviations 0.04--0.09, including level changes & 19 wavelengths, 443--660~nm; measured FWHM 1.37--3.20~nm. Modeled $G_{xy}=95\%$ of CIE 1964 spectral-locus area. Three-primary, power-limited calculation: largest $Y=1406~\mathrm{cd\,m^{-2}}$; gray-center $Y_{\max}=328~\mathrm{cd\,m^{-2}}$ (Figure~\ref{fig:gamut-luminance}) \\
\bottomrule
\end{tabularx}
\end{table}

For comparison in chromaticity and relative-luminance units, the same five 10-min series gave peak-to-peak ranges of 0.0011--0.0048 in $x_{10}$ and 0.0011--0.0029 in $y_{10}$. The standard deviation of luminance divided by its mean ranged from 0.21\% to 1.21\%. These values include the recorded level changes and describe the present CIE 1964 measurement protocol; they should not be equated with longer-term source stability or with repeatability of individual LED spectra.

The mean error of $1.25\,\Delta E_{00}$ reported by Machikhin et al. for 19 targets \cite{Machikhin2023} is similar in magnitude to the present per-color means of 0.34--1.26~$\Delta E_{00}$. A direct ranking is not justified: their study used the CIE 1931 observer and spatially averaged camera data, whereas the present measurements use central-field spectral radiance and the CIE 1964 $10^{\circ}$ observer. The target sets and sampling protocols also differ. The present study adds spatial color-difference maps and repeated central-color measurements under continuous feedback.

The LED and integrating-sphere studies offer complementary measurements of channel-spectral repeatability over short and six-month intervals \cite{Shi2024}, target-spectrum agreement for two stabilized illuminants \cite{Vernet2025}, and radiometric uniformity and stability over a large aperture \cite{Yuan2013}. The present homogenizer is substantially more compact than the meter-scale sphere arrangements in Table~\ref{tab:published-comparison} but retains a radiometric gradient across its 18-mm field. The comparison therefore concerns compactness and field uniformity, not a general performance ranking. Gamut conventions also differ: published primary triangles and qualitative descriptions of broad coverage are not equivalent to the present area ratio relative to the CIE 1964 spectral locus.

The AOC combines electronically selectable spectral lines a few nanometers wide, broad modeled chromaticity coverage, compact homogenization, and feedback regulation. Earlier work established AOTF color synthesis and proposed layouts for visual colorimetry \cite{Machikhin2023,Beliaeva2024Visual}. This study extends that work with spatial error maps, repeated measurements at five color centers, and separation of chromaticity and luminance mismatch. The stability measurement experiment also demonstrates regulated operation 12~h after calibration. The measurements establish how the instrument combines spectral flexibility with the field uniformity and reproducibility needed for controlled color-stimulus presentation.

\subsection{Routes to higher performance}

Controlled attenuation could improve effective drive resolution by distributing the available digital codes over the required line-radiance range while avoiding saturation. The 0.1\% command step specifies nominal digital resolution; colorimetric resolution also depends on the local slope and repeatability of the command--radiance response. Denser local calibration, inverse feed-forward compensation, and proportional--integral or gain-scheduled feedback could reduce residual error. These changes would require validation of settling time, disturbance rejection, noise amplification, and closed-loop bandwidth; none of these can be established from the 50-Hz update frequency alone.

Selecting primary wavelengths offers a complementary software approach. The calibration samples 19 wavelengths from 443 to 660~nm at 3-MHz intervals, whereas the driver supports 0.01-MHz frequency increments. Denser calibration would provide more candidate primary sets without changing the optics. For a specified target set, the software could select the smallest primary triangle enclosing the required chromaticities, subject to spectral separation, non-negative solutions, calibrated radiance ceilings, adequate conditioning of $\mathbf A_{\Lambda}(\mathbf c)$, and the RF-driver power limit. Figure~\ref{fig:gamut-luminance} shows why the selection must also provide sufficient luminance at the target chromaticity: a wide geometric gamut does not imply uniformly high output. Feasible candidates could then be ranked by the predicted color displacement for one 0.1\% command step to tailor resolution to the intended experiment.

Four-channel operation adds a degree of freedom but does not necessarily improve performance. The metameric family in permits optimization for quantization error, power margin, or observer agreement, but these criteria do not necessarily select the same spectrum. Validating four-channel operation therefore requires an explicit objective, calibration of cross-channel effects and radiance ceilings, and tests of the selected color and its neighborhood. The three-channel measurements reported here allow feedback and spatial mixing to be evaluated separately from this additional optimization problem.

\section{Conclusions}\label{sec:conclusions}

The AOTF-based colorimeter combines electronically tunable narrowband primaries, broad chromaticity coverage, compact spatial homogenization, and feedback regulation. Over the calibrated range of 443--660~nm, measured spectral-line FWHM values were 1.37--3.20~nm, and the calibrated primary set covered 95\% of the CIE 1964 $10^{\circ}$ spectral-locus area. Three-primary luminance mapping gave a calculated limit of $328~\mathrm{cd\,m^{-2}}$ at the gray Robertson chromaticity. The mean spatial error over an 18-mm field was $0.913\,\Delta E_{00}$, with residual non-uniformity dominated by luminance variation. With feedback enabled in three-channel operation, the five Robertson color centers gave mean central-field errors of 0.34--1.26~$\Delta E_{00}$ over 10~min each.

The combination of narrowband primaries and broad chromaticity coverage provides access to highly saturated color stimuli that are difficult to reproduce with conventional fixed-primary systems. Together with spatial homogenization and feedback-stabilized reproduction, this makes the colorimeter a platform for further color-vision experiments in highly saturated regions of color space, including studies of color discrimination, observer metamerism, and observer-dependent color matching.

\section*{Acknowledgments}
The authors thank Dr. K. V. Zotov of Soliton Photonics LLC for support with the supercontinuum source; A. B. Kozlov for support with the RF driver; and Dr. A. S. Machikhin for his comprehensive support of the project.
The research was carried out within the state assignment, approved by the Ministry of Science and Higher Education of the Russian Federation (theme No. FFNU-2025-0045).

\end{document}